\documentclass[showpacs,prl,twocolumn,floatfix]{revtex4}
\usepackage{graphicx,amssymb,amsmath}
\usepackage{esvect}
\usepackage[utf8]{inputenc}
\usepackage[T1]{fontenc}
\usepackage{amsmath,mathtools}
\usepackage{amssymb}
\usepackage{bm}
\usepackage{stmaryrd}

\newcommand{\overbar}[1]{\mkern 1.5mu\overline{\mkern-1.5mu#1\mkern-1.5mu}\mkern 1.5mu}

\begin{document}

\title{Betweenness centrality in dense spatial networks}

\author{Vincent Verbavatz}
\email{vincent.verbavatz@ipht.fr}
\affiliation{Institut de Physique Th\'{e}orique, CEA, CNRS-URA 2306, F-91191, 
Gif-sur-Yvette, France}
\affiliation{\'{E}cole des Ponts ParisTech, F-77420 Champs-sur-Marne, France}

\author{Marc Barthelemy}
\email{marc.barthelemy@ipht.fr}
\affiliation{Institut de Physique Th\'{e}orique, CEA, CNRS-URA 2306, F-91191, 
Gif-sur-Yvette, France}
\affiliation{CAMS (CNRS/EHESS) 54 Boulevard Raspail, 75006 Paris, France}

\begin{abstract}

The betweenness centrality (BC) is an important quantity for understanding the structure of complex large networks. However, its calculation is in general difficult and known in simple cases only. In particular, the BC has been exactly computed for graphs constructed over a set of $N$ points in the infinite density limit, displaying a universal behavior. We reconsider this calculation and propose an expansion for large and finite densities. We compute the lowest non-trivial order and show that it encodes how straight are shortest paths and is therefore non-universal and depends on the graph considered. We compare our analytical result to numerical simulations obtained for various graphs such as the minimum spanning tree, the nearest neighbor graph, the relative neighborhood graph, the random geometric graph, the Gabriel graph, or the Delaunay triangulation. We show that in most cases the agreement with our analytical result is excellent even for densities of points that are relatively low. This method and our results provide a framework for understanding and computing this important quantity in large spatial networks.  

\end{abstract}


\maketitle

\section{Introduction}

There are many centralities for characterizing the importance of a node in a network \cite{Rodrigues:2019}. Among those, local quantities fail to give interesting information about global structures while path-related measures are more relevant to describe the large-scale organization of networks. In particular, the betweenness centrality (BC), introduced in \cite{Freeman:1977} is a good probe of the structure of a network. Also, if one assumes that (i) individuals or goods travel on shortest paths in the network, and (ii) the demand is uniform (each pair of nodes constitutes an origin-destination couple) then the BC of a node (or an edge) corresponds to the local traffic that can be found at this node. In reality, the two assumptions are not always satisfied and how much of the real traffic the BC can explain is a debated question \cite{Holme:2003,Jayasinghe:2015, Kazerani:2009}. In general, highly congested points are signaled by very large values of the BC and this is relevant not only for transportation networks but also for communication networks such as the Internet where information packets can experience congestion problems at routers. In a router-based communication network, all nodes are connected to it directly and the BC is irrelevant in this case. A new direction for modern design of physical layer networks is to construct `wireless ad hoc networks' where routers are absent and packets of information are routed in a multihop fashion between any two nodes \cite{Santi:2003,Li:2009,Coon:2012}. This design allows for much larger and flexible networks and are nowadays realised under Wi-Fi direct standards. For these decentralized systems, the BC is a very relevant quantity and can be used as a criteria for identifying cluster nodes \cite{Gupta:2005}, or to identify the vulnerability backbone of the network \cite{Ercsey:2010}. Still in communication networks, it is intuitive to think that the traffic
between nodes tends to go through a small core of nodes. In this case,
the shortest paths are somehow curved inwards and it has been suggested
that this is related to the global curvature of the network
\cite{Narayan:2011, Jonckheere:2011}. A natural way to measure the
impact of the structure on the load in the network is then to
understand how the maximum traffic - approximated by the maximum BC - varies with various graph
properties and scales with the system size measured by the number of nodes  \cite{Narayan:2011}. 
Narayam and Saniee \cite{Narayan:2011} studied empirically various networks
and found essentially two families characterized by different values
of the exponent that governs this scaling. These authors proposed the idea that
this behavior is controlled by the curvature of the network and this
was justified mathematically by Jonckheere et
al. \cite{Jonckheere:2011}. 

The BC is of interest for spatial networks (planar or not) and real-world applications such as transport networks  
\cite{Lammer:2006,Crucitti:2006,Derrible:2012,Barthelemy:2013}. For street networks in cities, the study of street networks unveils the presence of crucial nodes with very large BC \cite{Lammer:2006,Strano:2012,Barthelemy:2013} and the localization of these congested points can reveal some interesting features about the organization of the network \cite{Strano:2012,Barthelemy:2013} and its large-scale organization \cite{Barthelemy:2004,Crucitti:2006,Barthelemy:2013}. More recently, an empirical study on almost 100 cities worldwide demonstrated that the empirical BC distribution seems to be an invariant in world cities \cite{Kirkley:2018} and that its structure results from the superimposition of a backbone tree (corresponding to the minimum spanning tree) and redundant streets in agreement with the picture proposed in \cite{Wu:2006} for synthetic networks. The interest in the BC also lies in the fact that it is correlated with some economic features such as the density of retail stores \cite{Strano:2009,Wang:2011,Porta:2012,Wang:2014,Davies:2015,Porta:2017} (we note that another analysis about Buenos Aires challenges this relation \cite{Scoppa:2015}). 

From a more theoretical point of view, however, few general results are known about the BC 
\cite{Barthelemy:2018,Gago:2012,Gago:2014}. Particular geometries such as branches and loops are well understood \cite{Lion:2017} while more involved geometries were studied only recently \cite{Lampo:2021}. Yet, a major theoretical result concerns planar graphs constructed on random points embedded in a bounded set. In this generic case, when there is an infinite number of points in the domain (usually a square or a disc), the shortest paths on the graph will likely be straight lines and the BC of any point can be computed exactly as a function of its position \cite{Giles:2015}. We note here that how the shortest path deviate from the straight line is interesting itself \cite{Aldous:2010} and more generally, the shape of shortest paths is an important problem \cite{Kartun:2019} and in relation with the first passage percolation problem, a well-known subject in statistical physics (see for example \cite{Auffinger:2017} and references therein). Here, we discuss to extend this infinite density calculation to the case of large but finite densities and the organization of this paper is as follows. We will first define the BC and recall some of its general properties and results (in particular for simple graphs). We will then present the perturbation expansion around the infinite density limit and test these results for various graphs constructed over a set of points in the plane.

\section{The betweenness centrality}

\subsection{Definition and generalities}

The betweenness centrality for a node $i$ in a graph $G$ with $N$ nodes is defined as \cite{Freeman:1977}
\begin{align}
  g(i)=\frac{1}{{\cal N}}\sum_{s\neq  t}\frac{\sigma_{st}(i)}{\sigma_{st}}
\end{align}
where $\sigma_{st}$ is the number of shortest paths from node $s$ to
node $t$ and $\sigma_{st}(i)$ the number of these shortest paths that
go through node $i$. The quantity ${\cal N}$ is a normalization that
we choose here  ${\cal N}=(N-1)(N-2)$ so that the BC is in $[0,1]$. We can define in a similar way the BC $g(e)$ for an edge $e$
using the quantity $\sigma_{st}(e)$ which is the number of shortest path from node $s$ to node $t$ going through the link $e$. 

It can be shown that the BC averaged over all nodes $\overline{g}=1/N\sum_ig(i)$ is proportional to the average shortest path $\ell$ \cite{Gago:2012,Barthelemy:2018}, which is shortest distance between two nodes in the graph, averaged over all pairs of nodes. This allows in particular to understand that adding a link to the graph will decreases the average BC. More precisely, it can be shown \cite{Gago:2012} that if we add to a graph of size $N$ a link of shortest path length $d$, we have 
\begin{align}
\overline{g}\rightarrow\overline{g}-\frac{2(d-1)}{N}
\end{align}
We note here that if the BC decreases on average, but this does not imply that the BC of all nodes decreases when adding new links. Locally, we can observe a increase of the BC of some points.

\subsection{One and two dimensional grids}

For one-dimensional lattices with $N$ nodes, it is easy to see that the BC of node $i$ 
($i\in \llbracket 1,N\rrbracket$) is given by
\begin{align}
g(i)=\frac{i}{N}\left(1-\frac{i}{N}\right)
\end{align}
The barycenter of all nodes $i_b=N/2$ is then also the most central node. There are other  results available in 1d and we mention here the example of the random geometric graph \cite{kartun:2021}.

\begin{figure}
\centering
\includegraphics[width=0.8\linewidth]{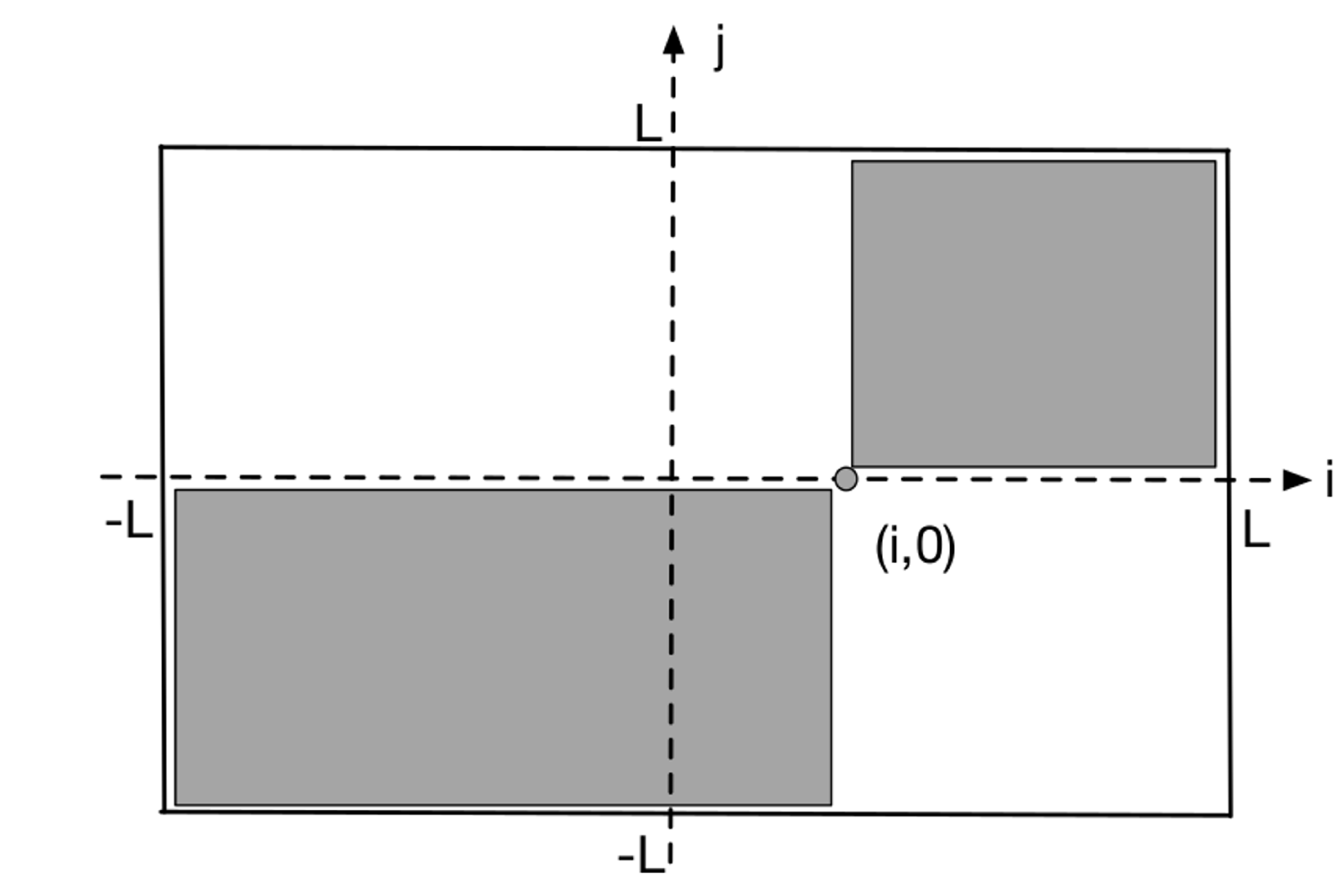}
\includegraphics[width=0.8\linewidth]{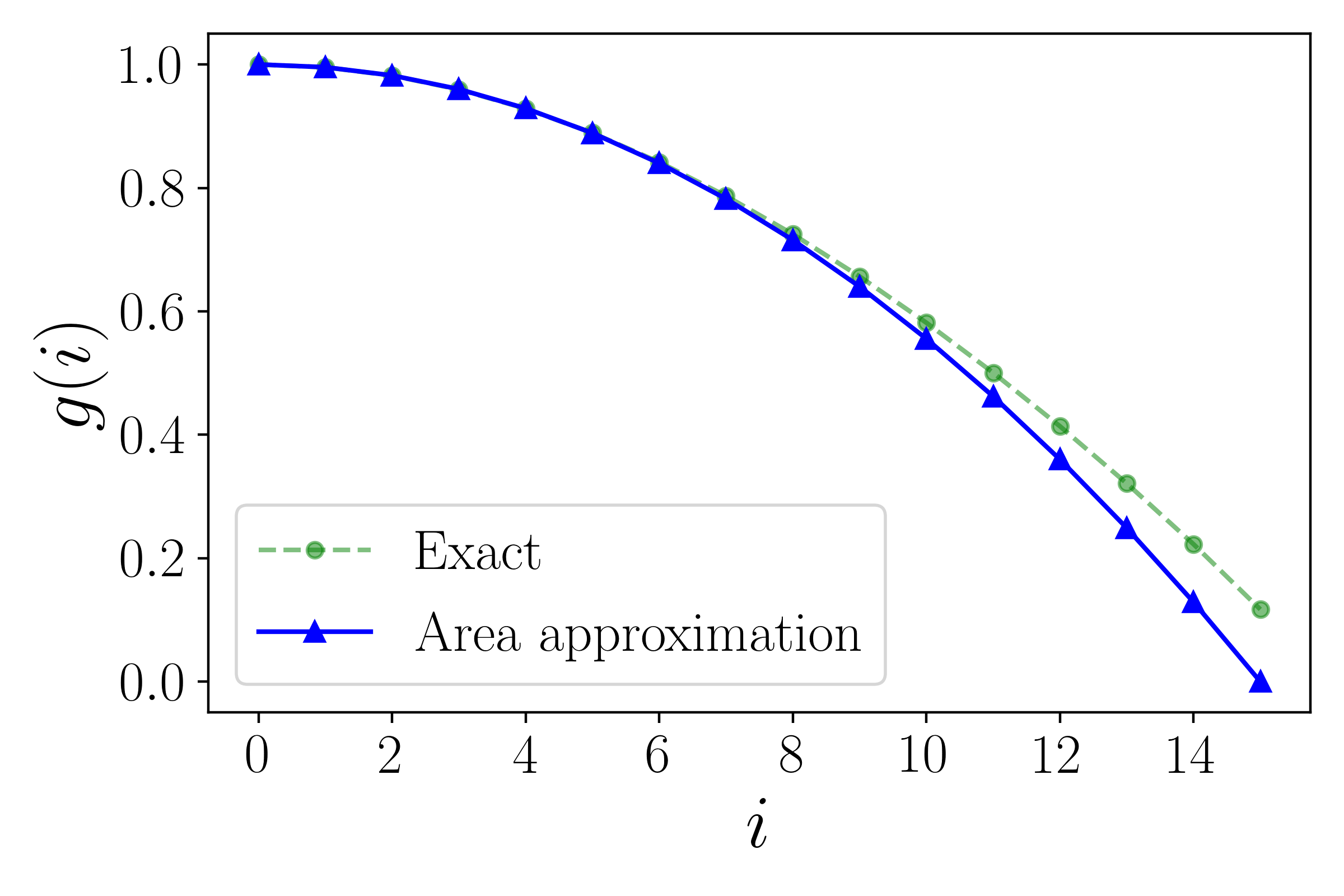}
\includegraphics[width=0.85\linewidth]{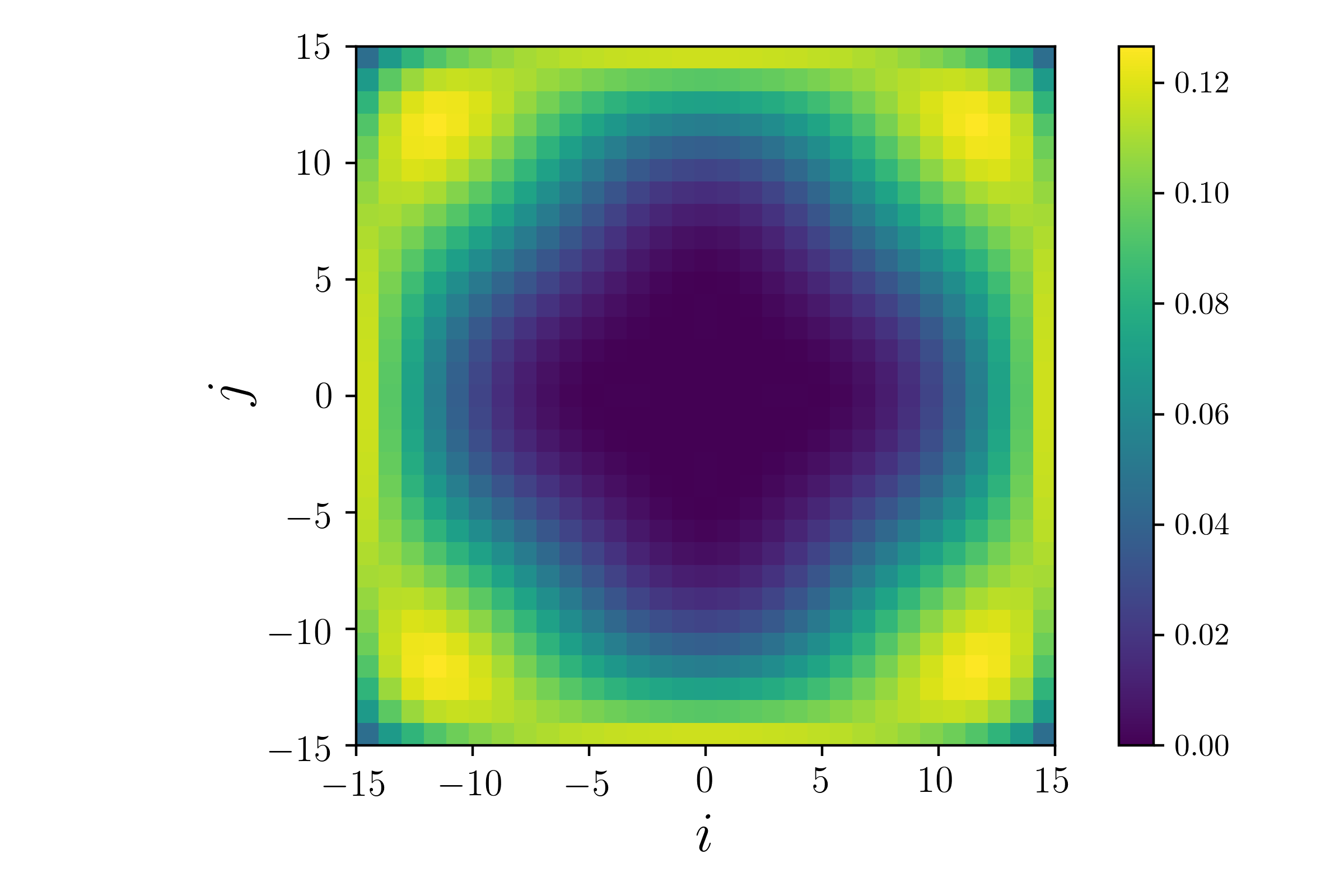}
\caption{(a) Approximation for computing the BC for the 2d grid: the BC at node $(i,0)$ is proportional to the product of shaded areas (up to a factor two accounting for the white squares).  (b) Comparison of the approximation of Eq. \ref{eq:appr} to the analytical formula of Eq. \ref{analytical-formula} computed numerically for $L=15$. (c) Relative error of the approximation of Eq. \ref{eq:appr_2d} on the square grid.
}
\label{fig:approx}
\end{figure}

For a two-dimensional square grid, it is easy to express the BC of a node 
as a sum of combinatorial factors that count the number of paths. The number of paths between points $(a,b)$ and $(i,j)$ with $a<i$ and $b<j$ being ${{i+j-a-b}\choose{i-a}}$, the centrality of the node $(i,j)$ on the grid 
$\llbracket -L,L\rrbracket\times \llbracket -L,L\rrbracket$ is:
\begin{align}
  \nonumber
  g(i,j)=\frac{1}{4L^2}&\sum_{\sigma \in \{-1,1\}}\sum_{a=-L}^{i}\sum_{b=-L}^{j}\\
  &\sum_{c=i}^L\sum_{d=j}^{L}
\frac
{ {{j-b+\sigma(i-a)}\choose{\sigma(i-a)}} {{d-j+\sigma(c-i)}\choose{\sigma(c-i)}} }
{ {{c+d-a-b}\choose{\sigma(c-a)}} }
\label{analytical-formula}
\end{align}
where $\sigma=\pm 1$ corresponds to nodes with $j<0$ and $j>0$, respectively. 
This expression is difficult to analyze, but we can resort to the simple approximation described in Fig.~\ref{fig:approx}(a).
We assume that the number of paths going through the node $(i,0)$ is proportional to the product of areas described in gray in Fig.~\ref{fig:approx}, normalized by the total number of paths (we have to multiply the result by a factor $2$ by symmetry). We thus obtain
\begin{align}
  \nonumber
g(i, 0)&\propto L(L-i)\times (i+L)L/L^4\\
       &=1-\left(\frac{i}{L}\right)^2
         \label{eq:appr}
\end{align}
We compare this approximation to the exact numerical result showing a very good agreement (Fig.~\ref{fig:approx}(b)). The discrepancy appears essentially for $i\approx L$ where the approximation predicts $g(L)=0$ which is exact for $L\gg 1$. Using this same argument, we find that for any node located at $(i,j)$, we find
\begin{equation}
  g(i,j) \simeq (1-i^2)(1-j^2)
  \label{eq:appr_2d}
\end{equation}
We compare this expression to the exact result and show in Fig.~\ref{fig:approx} the relative error. Here also the most important errors appear at the boundary of the square but, despite its simplicity, the approximation is very good in the bulk of the square.

\subsection{Loops, branches, and more complex graphs}

More complex graphs have also been studied from the perspective of the BC. In particular, in \cite{Lion:2017} a toy model made of a star network with $N_b$ branches of size $n$ and links of weight $1$, superimposed to a loop at distance $\ell$ from the center and with links of weight $w$ was considered. The BC at the center and on the loop were computed and it can be shown that the loop can be more central than the center if $w<w_c$ where the threshold scales as $w_c\sim n/N_b$. This sheds some light on empirical results about road networks where ring roads can be more central than the spatial center of the system.

Also recently a more complex structure was considered in \cite{Lampo:2021}. In this study, the authors introduce a family of planar graphs composed by a square grid connected to an arboreal periphery.  The BC was then computed at the center of the grid and some other important points such as the ones connecting the square grid to the peripheral trees. 

In general, for complex networks that are not planar, the BC is increasing with the degree of the node as a power law $g(k)\sim k^\eta$ with an exponent $\eta$ that is in general less than $2$. The main reason for this bound is that the
number of possible paths through a node of degree $k$ is
$k(k-1)$ which scales as $k^2$ for large $k$. This is then what we would obtain for the BC
if all neighbors lead to regions with roughly the same number of nodes. If it is
not the case, some neighbors will be more important than others and then
not all the $k(k-1)$ paths are important and therefore $\eta<2$.

\subsection{Distribution of the BC in planar graphs}

The distribution of the BC $g$ was discussed recently in \cite{Kirkley:2018} where it was shown on almost 100 different road networks for worldwide cities that its probability distribution $P(g)$ of having a certain BC $g$ is invariant. This invariance is a consequence of a bimodal regime where the high BC nodes belong to the underlying tree structure of the graph, and the low BC nodes to loops that provide alternate paths. If we rescale the centrality by the number $N$ of nodes $\tilde{g}=g/N$, we  obtain the invariant distribution as
\begin{align}
P(\tilde{g})\sim\frac{\mathrm{e}^{-\tilde{g}/\beta}}{\tilde{g}^\alpha}
\end{align}
where the exponent $\alpha\approx 1$ can be explained with a simple tree model \cite{Kirkley:2018}, and where $\beta$ depends on the specific graph. This invariance in particular suggests that the interesting information about the BC lies not in its statistical properties but rather in its spatial distribution, and where the high BC nodes are located \cite{Barthelemy:2013,Kirkley:2018} which depends in general on details of the structure of the graph.

\section{Betweenness centrality in dense and quasi-dense graphs}

\subsection{Graphs constructed over a set of points}

We will consider here different graphs constructed over a set of points in a bounded domain. More precisely, we consider a Poisson process where $N$ points are distributed randomly in a plane domain $\mathcal{D}$ of area $V$ (which will be a disk or a square). The density of nodes is denoted by $\rho=N/V$. There are multiple ways to connect these points to each other and we will consider here various graphs. 

First, we will consider graphs that are constructed by connecting a node to its $k$-nearest neighbors (\textbf{$k$-NN} with $k=7$), the random geometric graph (\textbf{RGG}) that connects points closer than a threshold distance $d$ (we choose $d=2/\sqrt{\rho}$), the minimal spanning tree (\textbf{MST}) that connects all the vertices together, without any cycles and with the minimum possible total edge weight, the Delaunay triangulation (\textbf{DT}) that gives a triangulation such that no point is inside the circumcircle of any triangle of the triangulation, the Grabriel Graph (\textbf{GG}) which is the subgraph of DT where any two distinct points $P$ and $Q$ are adjacent precisely when the closed disc having $PQ$ as a diameter contains no other points and finally the relative neighborhood graph (\textbf{RNG}) which connects two points $P$ and $Q$ by an edge whenever there does not exist a third point $R$ that is closer to both $P$ and $Q$ than they are to each other. These graphs represent many important cases and are widely studied. Understanding the BC for these cases thus represents an important step towards a general theory of the BC in spatial networks.

\subsection{Perturbation around the infinite density}

We compute the BC of nodes in $\mathcal{D}$ in the quasi-dense limit ($1 \ll \rho < \infty$). We aim at finding an expression for the BC that depends only on the absolute position of points in the plane and not the specific graph. In \cite{Giles:2015}, it was shown that in the dense limit ($\rho = \infty$) on a disk, the BC of a node depends only on its distance to the center, whatever the specific graph. This approximation relies on the fact that shortest paths in this limit are essentially straight lines, which explains the universality of the result. On the other hand, for finite densities, the shortest paths display significant transversal deviations and we expect non-universal corrections. 

We present here a perturbation expansion at the lowest non-trivial order of this previous result when the density is finite. We denote by $(i, j, \kappa)$ random nodes (among $N$ nodes of a graph $G$) inside a disk domain $\mathcal{D}$ of area $V$. The quantity $SP(i,j)=\left\{x_{ij_1}, ..., x_{ij_m}, ..., x_{ij_n}\right\}$ denotes the shortest path between points $i$ and $j$ (that we assume to be unique, which for spatial networks is expected - a degeneracy would imply exactly the same distance between two nodes which is very unlikely, in contrast with the topological distance which is an integer that counts the number of jumps). 

We define the indicator function
\begin{align}
  \sigma_{ij}(\kappa) = \mathbf{1}_{\kappa \in SP(i,j)} = \sum_m  \mathbf{1}_{\kappa = x_{ij_m}}
\end{align}
This indicator function $\sigma_{ij}(\kappa)$ is equal to unity if $\kappa$ is in $SP(i,j)$ and zero otherwise.
The betweenness centrality for the node $\kappa$ is then
\begin{equation}
g(\kappa)=\frac{1}{2}\sum_i \sum_j \frac{\sigma_{ij}(\kappa)}{\sigma_{ij}}
\end{equation}
where $i$ and $j$ are nodes of the graph. For large $\rho$, we use a continuous approximation and write
\begin{equation}
\sigma_{ij}(\kappa) = \int_{t_i}^{t_j} \mathrm{d}t~\delta(\bm{\kappa} - \bm x(t))
\end{equation}
where the shortest path $\{ \bm x(t)\in SP(i,j) \}$ is parametrized by $t\in [t_1,t_2]$ where $t_1$ and $t_2$ correspond to the endpoints ($\delta$ is the Dirac delta function). The BC for $\kappa$ is then
\begin{equation}
g(\kappa) = \frac{1}{2V^2}\int_{\mathcal{D}} \mathrm{d}\mathbf{r}_i \int_{\mathcal{D}} \mathrm{d}\mathbf{r}_j~\sigma_{ij} (\kappa)
\end{equation}

In the continuous limit \cite{Giles:2015} $\rho \rightarrow \infty$, the shortest paths are straight segments and the indicator reads as $\sigma_{ij}=\delta(x\cos(\phi) +y\sin(\phi)-p)$ where $\kappa=(x,y)$ and where the segment $(i,j)$ is parametrized by $p$ and $\phi$ (for this type of parametrization, see for example \cite{Santalo:2004}). This means that $\kappa$ is in $SP(i,j)$ if and only if $\kappa$ is on the line between $i$ and $j$. In particular, it is easy to check that $\int_{\mathcal{D}}\mathrm{d}\bm\kappa~\sigma_{ij}(\kappa) = |t_2-t_1|$ as expected.

\begin{figure}
\centering
\includegraphics[width=0.3\textwidth]{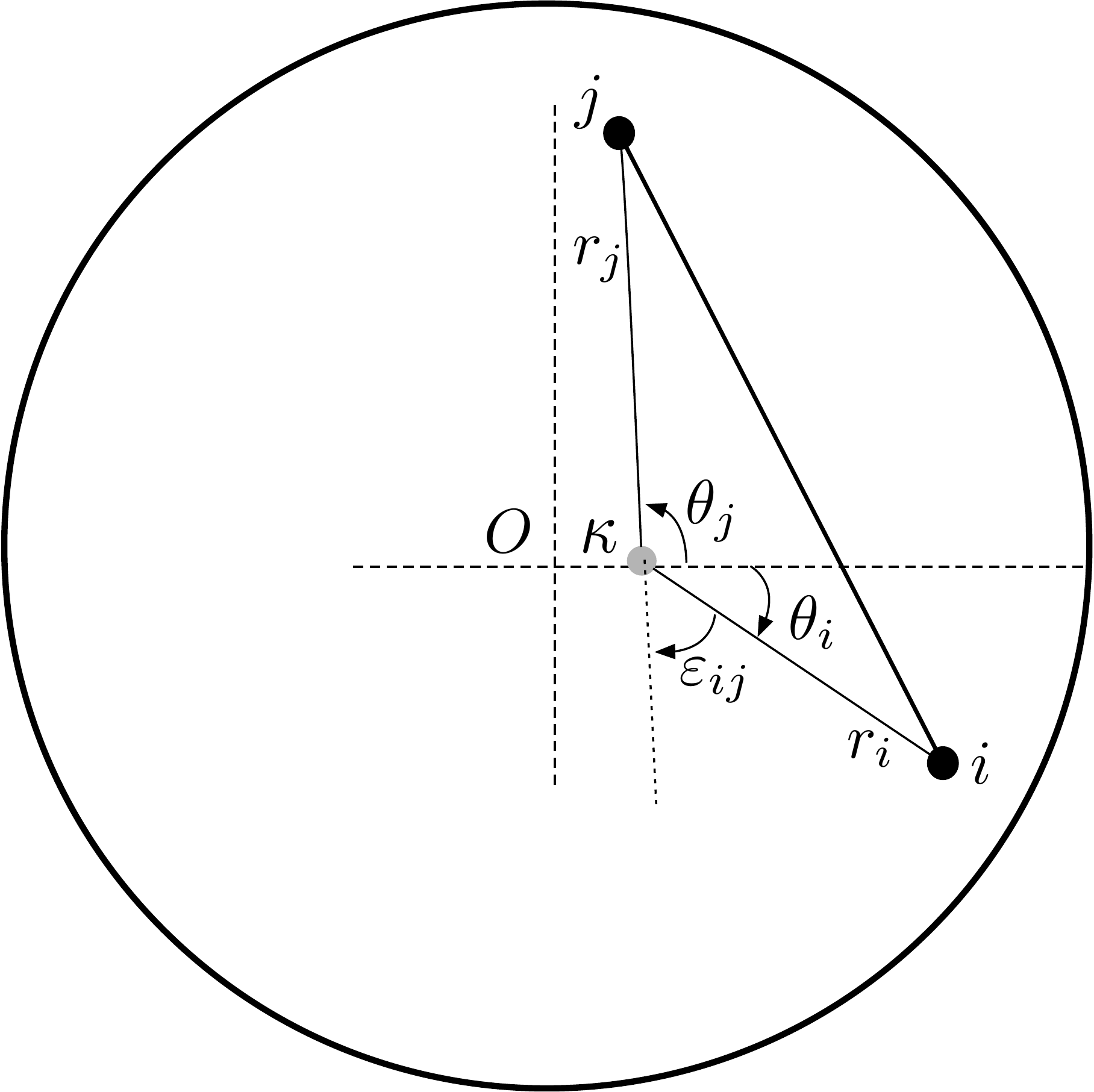}
\caption{Sketch of the system considered and notations. The origin of the polar system is $\kappa$ and the nodes $i$ and $j$ have the coordinates $(r_i,\theta_i)$ and $(r_j,\theta_j)$ in this system. The deviation of the two segments $(i\,\kappa)$ and $(\kappa, j)$ from the straight line is characterized by the angle $\varepsilon_{ij}=\theta_i-\theta_j+\pi$.}
\label{fig:kartun}
\end{figure}

In the quasi-dense limit ($1 \ll \rho < \infty$), we define the average betweenness centrality for $\kappa$ as the expectation the BC for $\kappa$
\begin{equation}
\overbar{g(\kappa)}=\mathbb{E}_G(g(\kappa))
\end{equation}
where $\mathbb{E}_G$ denotes the average over all the graphs (for a given connection rule) constructed over an ensemble of points realizations. 
We then obtain
\begin{eqnarray}
  \overbar{g(\kappa)}=\frac{1}{2V^2}\int_{\mathcal{D}}
  \mathrm{d}\mathbf{r}_i \int_{\mathcal{D}} \mathrm{d}\mathbf{r}_j~\mathbb{E}_G(\sigma_{ij} (\kappa))
\end{eqnarray}
The quantity $\sigma_{ij}(\kappa)$ is an indicator function and its average is then a probability
\begin{align}
\mathbb{E}_G(\sigma_{ij} (\kappa))= \mathrm{Prob}\left(\kappa \in SP(i,j)\right)
\end{align}
which we will denote by $\chi_{ij}(\kappa)=\mathrm{Prob}\left(\kappa \in SP(i,j)\right)$.
In the dense limit, shortest paths are straight lines and we have
\begin{align}
  \chi_{ij}(\kappa)=\delta(x\cos(\phi) +y\sin(\phi)-p)
\end{align}
When the density is finite, the shortest paths deviate from the straight line and we define the angular deviation $\epsilon_{ij}$ from the straight line $(i,j)$ in the frame of origin $\kappa$:  $ \epsilon_{ij} = \theta_i - \theta_j + \pi $ (see Fig.~\ref{fig:kartun}). Due to the statistical isotropy of the problem, it is enough to consider the node at distance $\kappa$ from the center and at polar angle $\theta=0$ and here and in the following we will work with the polar coordinate centered on this point. 

We now express the probability $\chi_{ij} (\kappa)$ that $\kappa$ is in $SP(i,j)$ for a given value of $\epsilon_{ij}$, and the average BC can then formally be rewritten as the 5d integral
\begin{align}
  \nonumber
  \overbar{g(\kappa)} = &\frac{1}{2V^2}\int_{\mathcal{D}^2} \mathrm{d}\mathbf{r}_i\mathrm{d}\mathbf{r}_j\\
  &\int_0^{\infty} \mathrm{d} \epsilon_{ij} \delta(\theta_i - \theta_j + \pi - \epsilon_{ij}) \chi_{ij} (\kappa | \epsilon_{ij})
\end{align}
where $\chi_{ij} (\kappa | \epsilon_{ij})$ is the probability that $\left\{\kappa \in SP(i,j)\right\}$ conditioned by $\epsilon_{ij}$. The delta function $\delta(\theta_i - \theta_j + \pi - \epsilon_{ij})$ ensures the definition of the angle $\epsilon_{ij}$.

\begin{figure}
\centering
\includegraphics[width=0.5\textwidth]{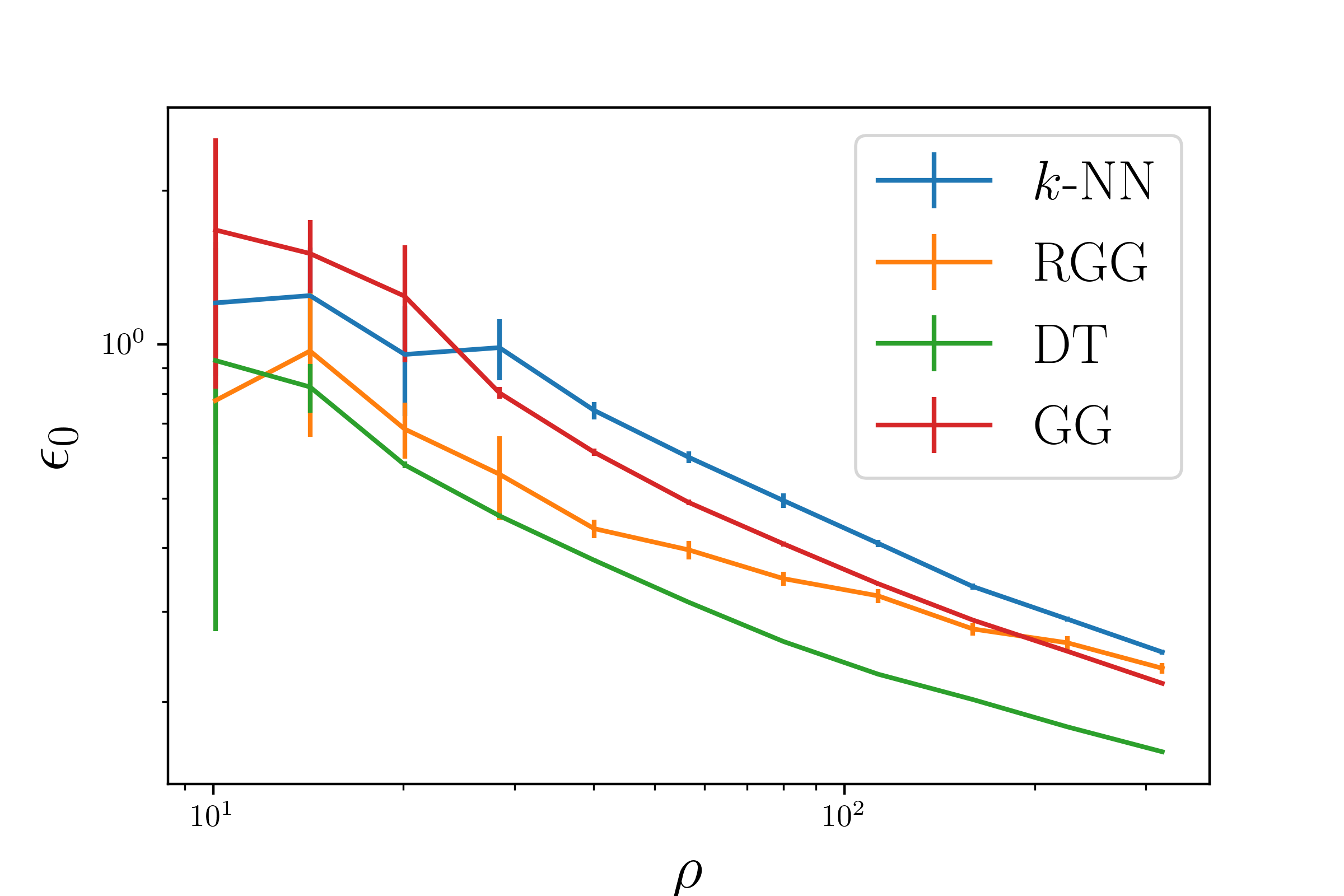}
\caption{Writing $\chi(\epsilon)=\epsilon_0(\rho)\mathcal{N}e^{-\epsilon/\epsilon_0(\rho)}$, we show that $\epsilon_0$ is a smooth decreasing function of the density $\rho$, validating the shape of $\chi(\epsilon)$ for $k$-NN, RGG, DT and GG graphs (the vertical error bars correspond to the dispersion). The graph suggests a power law relation of the form $\epsilon_0 (\rho) \simeq \rho^{-\beta}$ with $\beta \simeq 0.4 \pm 0.1$. We note that the exponent is not the exact same for all graphs, nor is the prefactor, thus making some types of graphs converging faster towards the dense regime than others.}
\label{fig:exp}
\end{figure}

In the infinite density limit, we know from \cite{Giles:2015} that this conditional probability is given by
\begin{align}
  \chi_{ij} (\kappa | \epsilon_{ij},\rho=\infty) \propto \left(\frac{1}{r_i} + \frac{1}{r_j}\right) \delta(\epsilon_{ij})
\end{align}
and motivated by this case we assume the following generalization
\begin{align}
  \chi_{ij} (\kappa | \epsilon_{ij}) = \left(\frac{1}{r_i} + \frac{1}{r_j}\right) \chi(\epsilon_{ij})
  \label{eq:chi}
\end{align}
where $\chi(\epsilon)$ is an unknown function. Denoting $\epsilon_{ij}$ by $\epsilon$, we assume that  $\chi(\epsilon)$ is independent of $(i,j, \kappa)$. This is a strong assumption that we empirically show to be correct for the $k$-NN, RGG, DT and GG graphs (as we will further discuss below, this approximation is incorrect in the two cases of the MST and the RNG graphs and suggests that for these graphs the assumption that $\chi(\epsilon)$ is independent of $(i,j, \kappa)$ is not correct). Indeed, we show empirically that for these graphs, the function
\begin{align}
  \chi(\epsilon) = \frac{\chi_{ij} (\kappa | \epsilon)}{ \left(\frac{1}{r_i} + \frac{1}{r_j}\right)}
\end{align}
can be fitted by a decreasing exponential function of $\epsilon$ with parameter $\epsilon_0$ ($R^2>0.8$ for all graphs except MST and RNG)
\begin{align}
  \chi(\epsilon)=\epsilon_0(\rho)\mathcal{N}e^{-\epsilon/\epsilon_0(\rho)}
\end{align}
where $\epsilon_0(\rho)$ is a smooth decreasing function of $\rho$ (see Fig. \ref{fig:exp}) and $\mathcal{N}$ a normalization.

Using this form Eq.~\ref{eq:chi}, we obtain
\begin{align}
\nonumber
\overbar{g(\kappa)} = \frac{1}{2V^2}\int_0^{\infty}\mathrm{d} \epsilon~\chi(\epsilon)\int_{\mathcal{D}} \mathrm{d}\mathbf{r}_i \int_{\mathcal{D}} \mathrm{d}\mathbf{r}_j \\
\left(\frac{1}{r_i} + \frac{1}{r_j}\right)\delta\left(\theta_i - \theta_j + \pi -  \epsilon\right) 
\label{eq:aveg}
\end{align}
In the dense limit, we have $\chi(\epsilon)=\mathcal{N}\delta(\epsilon)$ and we recover the known result of \cite{Giles:2015}. In order to go beyond this infinite density result, we expand this function $\chi(\epsilon)$ around 0 to the second order in $\epsilon_o$
\begin{align}
\nonumber
\chi(\epsilon)&=\epsilon_0(\rho)\mathcal{N}e^{-\epsilon/\epsilon_0(\rho)}\\
&\simeq\mathcal{N}\left(\delta(\epsilon) - \epsilon_0(\rho)\delta'(\epsilon) +  \epsilon_0^2(\rho)\delta''(\epsilon)\right)
\label{first-exp}
\end{align}
Here, we used the distributional derivative of the Dirac delta function, which is defined so that for any compactly supported smooth test function $\phi$, we have
\begin{equation}
\int \mathrm{d}x \phi(x)\delta'(x) = - \int \mathrm{d}x \phi'(x)\delta(x)
\end{equation}

Inserting the expansion of Eq. \ref{first-exp} into the expression Eq.~\ref{eq:aveg}, we get
\begin{align*}
\nonumber
\overbar{g(\kappa)} &= \frac{\mathcal{N}}{2V^2}\int_0^{\infty} \mathrm{d}\epsilon~\left(\delta(\epsilon) - \epsilon_0(\rho)\delta'(\epsilon) +  \epsilon_0^2(\rho)\delta''(\epsilon)\right)\\
& \int_{\mathcal{D}} \mathrm{d}\mathbf{r}_i \int_{\mathcal{D}} \mathrm{d}\mathbf{r}_j ~\left(\frac{1}{r_i} + \frac{1}{r_j}\right)\delta\left(\theta_i - \theta_j + \pi -  \epsilon\right) \\
\nonumber
&= \frac{\mathcal{N}}{4V^2} \int_{0}^{2\pi}\mathrm{d}\theta_i~r(\theta_i)r(\theta_i + \pi)\left(r(\theta_i) + r(\theta_i + \pi)\right) \\
&- \epsilon_0(\rho) \int_{0}^{2\pi}\mathrm{d}\theta_i~r(\theta_i)r'(\theta_i + \pi)\left(r(\theta_i) + 2r(\theta_i + \pi)\right)  \\
\nonumber
 &- \epsilon_0^2(\rho) \int_{0}^{2\pi}\mathrm{d}\theta_i~r(\theta_i)r''(\theta_i + \pi)\left(r(\theta_i) + 2r(\theta_i + \pi)\right) 
 \end{align*}
 In the polar coordinate centered at $\kappa$, the frontier of the disc is given by $r(\theta) = \sqrt{R^2-\kappa^2 \sin^2(\theta)} - \kappa \cos(\theta)$ and the previous expressions can be now rewritten as
\begin{align}
\nonumber
&\overbar{g(\kappa)} 
= \frac{\mathcal{N}}{4V^2}
\Big[
2\int_{0}^{2\pi}\mathrm{d}\theta~(R^2 - \kappa^2) \sqrt{R^2 - \kappa^2 \sin^2(\theta)} \\ 
&-\epsilon_0^2(\rho) \int_{0}^{2 \pi} \mathrm{d}\theta~
\frac{
\kappa^{2} (R^{2} - \kappa^{2}) (R^{2} (6 \sin^{2}\theta - 1) - 5 \kappa^{2} \sin^{4}\theta)}
{(R^{2} - \kappa^{2} \sin^{2}\theta)^{3/2}}
\Big]
\end{align}

The first order term (coefficient of $\epsilon_0$)  is equal to 0 and the first non-trivial term is of second order. We introduce the functions
\begin{align}
I_0(\kappa, R)=2\int_{0}^{2\pi}\mathrm{d}\theta~(R^2 - \kappa^2) \sqrt{R^2 - \kappa^2\sin^2(\theta)}
\end{align}
and 
\begin{align}
\nonumber
&I_2(\kappa, R)=\\
&\int_{0}^{2 \pi} \mathrm{d}\theta~\frac{\kappa^{2} \left(R^{2} - \kappa^{2}\right) \left(R^{2} \left(6 \sin^{2}{\left(\theta \right)} - 1\right) - 5 \kappa^{2} \sin^{4}{\left(\theta \right)}\right)}{\left(R^{2} - \kappa^{2} \sin^{2}{\left(\theta \right)}\right)^{3/2}}
\end{align}
and the BC can be rewritten as 
\begin{align}
\overbar{g(\kappa)} 
  = \frac{\mathcal{N}}{4V^2}[I_0(\kappa)-\epsilon_0^2(\rho)I_2(\kappa,R)]
  \label{eq:resfin}
\end{align}
In the result of Eq. \ref{eq:resfin}, the infinite density limit which corresponds to the first term is universal, i.e. independent from the graph structure. In contrast, the second term (term
in $\epsilon_0(\rho)$) does depend on the graph and encodes the deviation of shortest paths from the straight line which varies from a graph to another. This implies that in general (and as expected) the BC of a graph at finite density is not universal and depends on the graph considered. In particular, the numerical result of Fig. \ref{fig:exp} suggests a power law relation of the form $\epsilon_0 (\rho) \simeq \rho^{-\beta}$ with $\beta \simeq 0.4 \pm 0.1$.  We observe that the value of the exponent $\beta$ and the prefactor are not exactly the same for all graphs, implying different rates of convergence towards the dense regime.

We can express the  integrals appearing in Eq.~\ref{eq:resfin} using special functions
and we get
\begin{equation}
I_0(R, \kappa)= 2R(R^2-\kappa^2)E\left(\frac{\kappa}{R}\right)
\label{eq:infinite}
\end{equation}
and
\begin{align}
I_2(R, \kappa)= 8R^3  \left[ 3 \left(\frac{\kappa}{R}\right)^2 K\left(\frac{\kappa}{R}\right) + 2 K\left(\frac{\kappa}{R}\right) \right. \\ \left. + 2\left(\frac{\kappa}{R}\right)^2E\left(\frac{\kappa}{R}\right) - 2 E\left(\frac{\kappa}{R}\right) \right]
\end{align}
where $K(x)$ and $E(x)$ are respectively the elliptic integrals of the first and second kind.

If $\kappa \ll R$, we can show that
\begin{align}
I_2(R, \kappa)= 4\pi R^3\left(\frac{\kappa}{R}\right)^2 + o\left(\left(\frac{\kappa}{R}\right)^2\right)
\end{align}
while if $R - \kappa \ll R$
\begin{align}
I_2(R, \kappa) \simeq 16 R^3\left(1-\frac{\kappa}{R}\right)\log\left(1-\frac{\kappa}{R}\right) 
\end{align}

Normalizing the BC by $\overbar{g(0)}=\frac{\mathcal{N}}{4V^2} \times 8\pi R^3$, we obtain
\begin{equation}
g^*(\kappa) = \frac{\overbar{g(\kappa)}}{\overbar{g(0)}} =
\frac{1}{8\pi R^3}[I_0(\kappa)-\epsilon_0^2(\rho)I_2(\kappa,R)]
\label{eq:final}
\end{equation}
If $\kappa \ll R$, it gives
\begin{align}
g^*(\kappa) \simeq 1 - \left(5 + \frac{1}{2}\epsilon_0^2(\rho) \right)\left(\frac{\kappa}{R}\right)^2
\end{align}
while if $R - \kappa \ll R$
\begin{align}
g^*(\kappa) \simeq \left(\frac{4}{\pi} - 2 \epsilon_0^2(\rho)\log\left(1-\frac{\kappa}{R}\right)\right)\left(1-\frac{\kappa}{R}\right)
\end{align}

We note that under such a normalization we have always $g^*(0) = 1$ and $g^*(R) = 0$ (which can be easily proven). In order to use this result and to compare it with the simulations, we need to specify the deviation characterized by $\epsilon_0(\rho)$ (see below for the numerical study).


\subsection{Numerical study}

\begin{figure}
  \centering
  \includegraphics[width=0.5\textwidth]{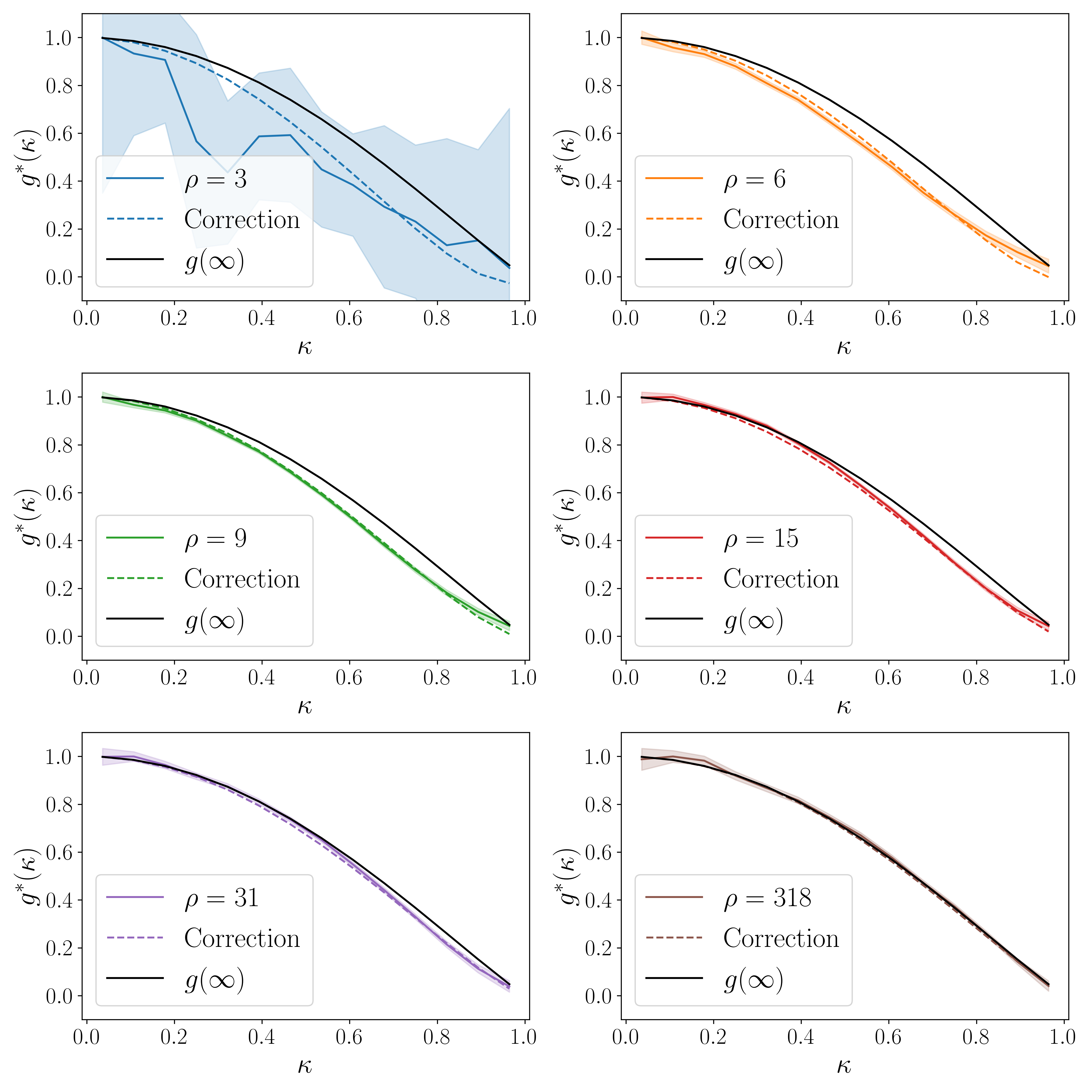}
  \caption{Comparison of the expansion of Eq. \ref{eq:final} with the numerical result for the DT.  The quality of the approximation increases with the density but is insightful at surprisingly low densities (6 points per square unit is less than 20 points in the disc). The number of points is $N = \rho \pi$.}
\label{fig:DT}
\end{figure}

\begin{figure}
\centering
\includegraphics[width=0.5\textwidth]{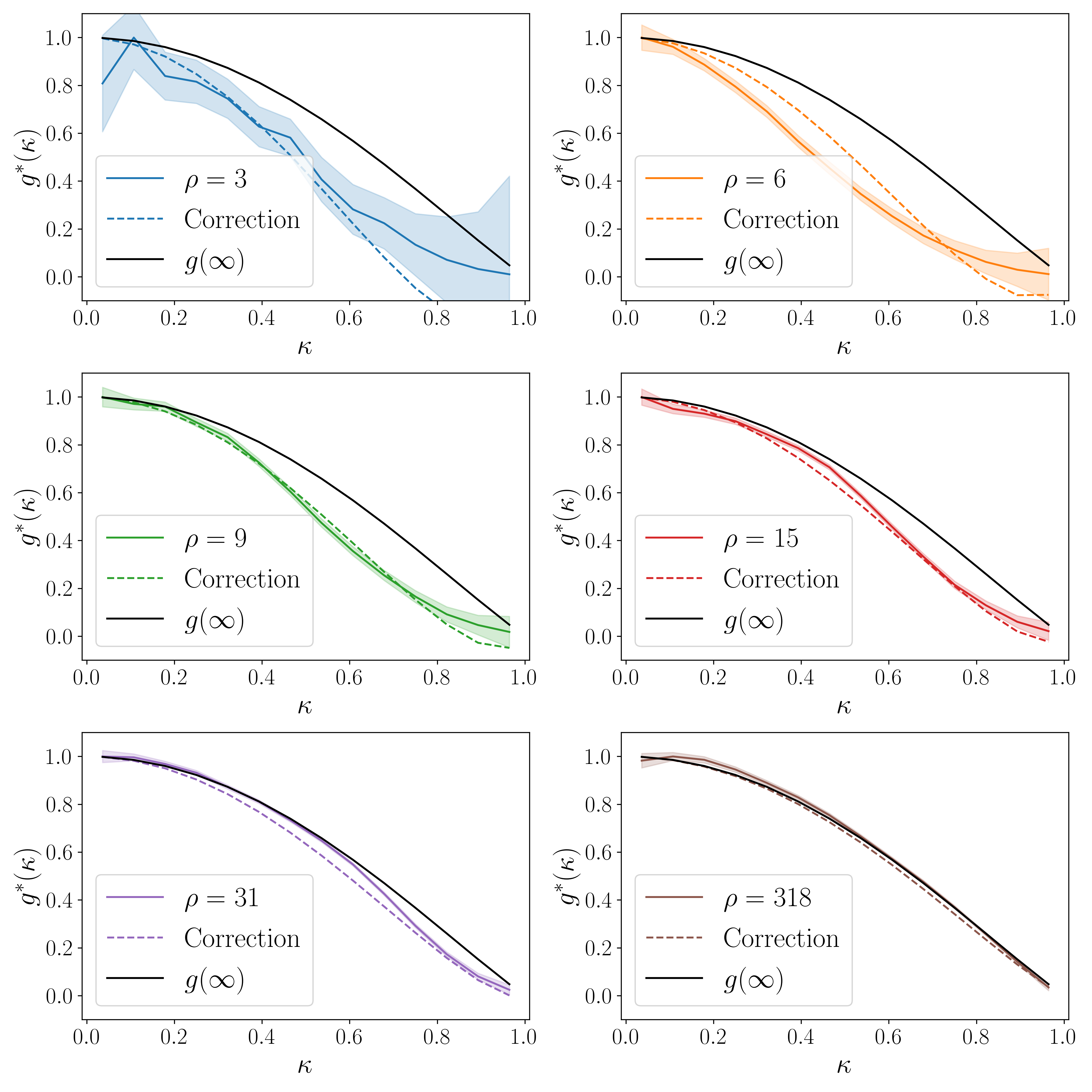}
\caption{Comparison of the expansion of Eq. \ref{eq:final} with the numerical result for the $k$-NN.  The quality of the approximation increases with the density but it is already very good for a density $\rho=9$ or larger. The number of points is $N = \rho \pi$.}
\label{fig:kNN}
\end{figure}

\begin{figure}
\centering
\includegraphics[width=0.5\textwidth]{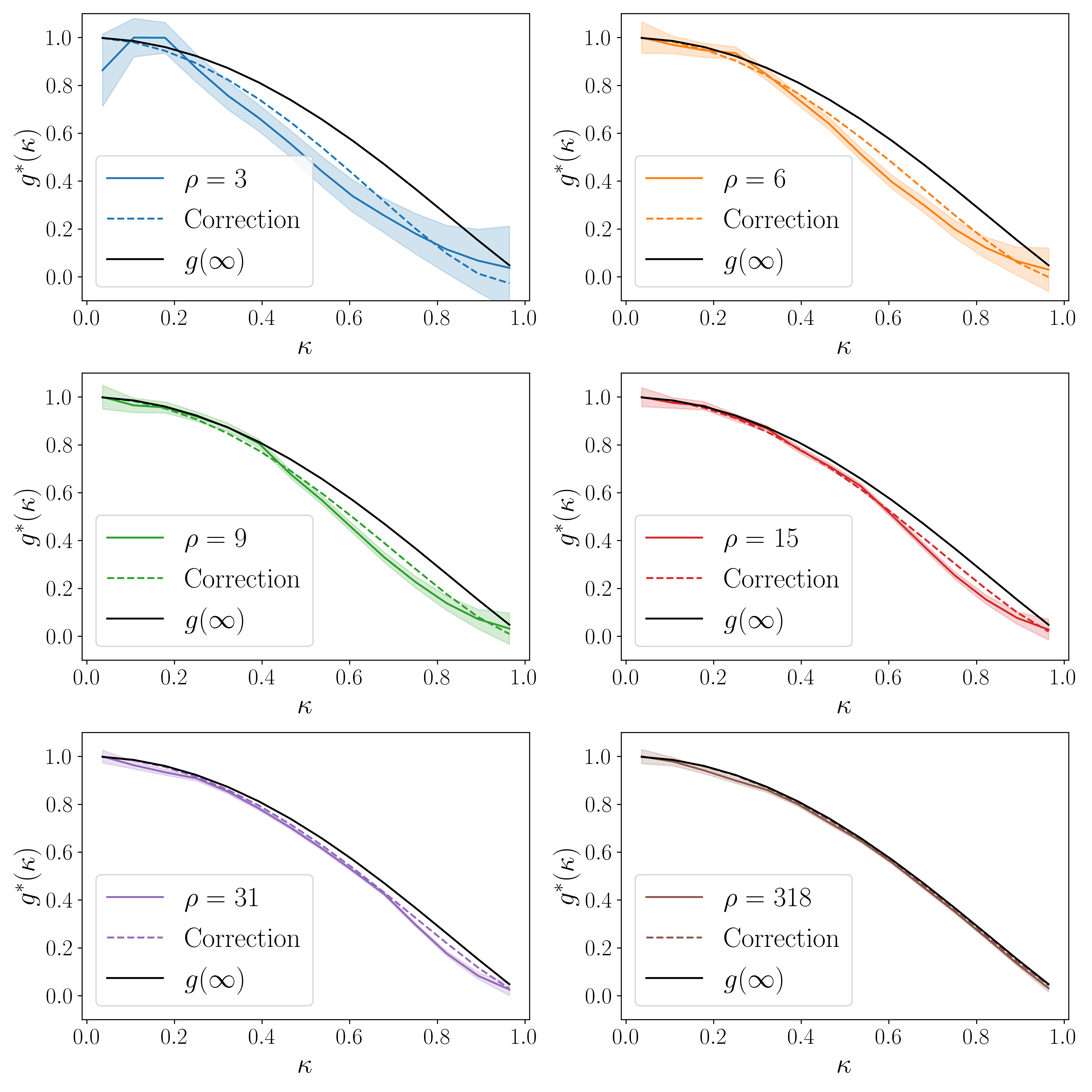}
\caption{Comparison of the expansion of Eq. \ref{eq:final} with the numerical result for the RGG. The convergence to the infinite density limit is faster than for other graphs and the approximation is very good for densities as low as $\rho=9$. The number of points is $N = \rho \pi$.}
\label{fig:RGG}
\end{figure}

\begin{figure}
\centering
\includegraphics[width=0.5\textwidth]{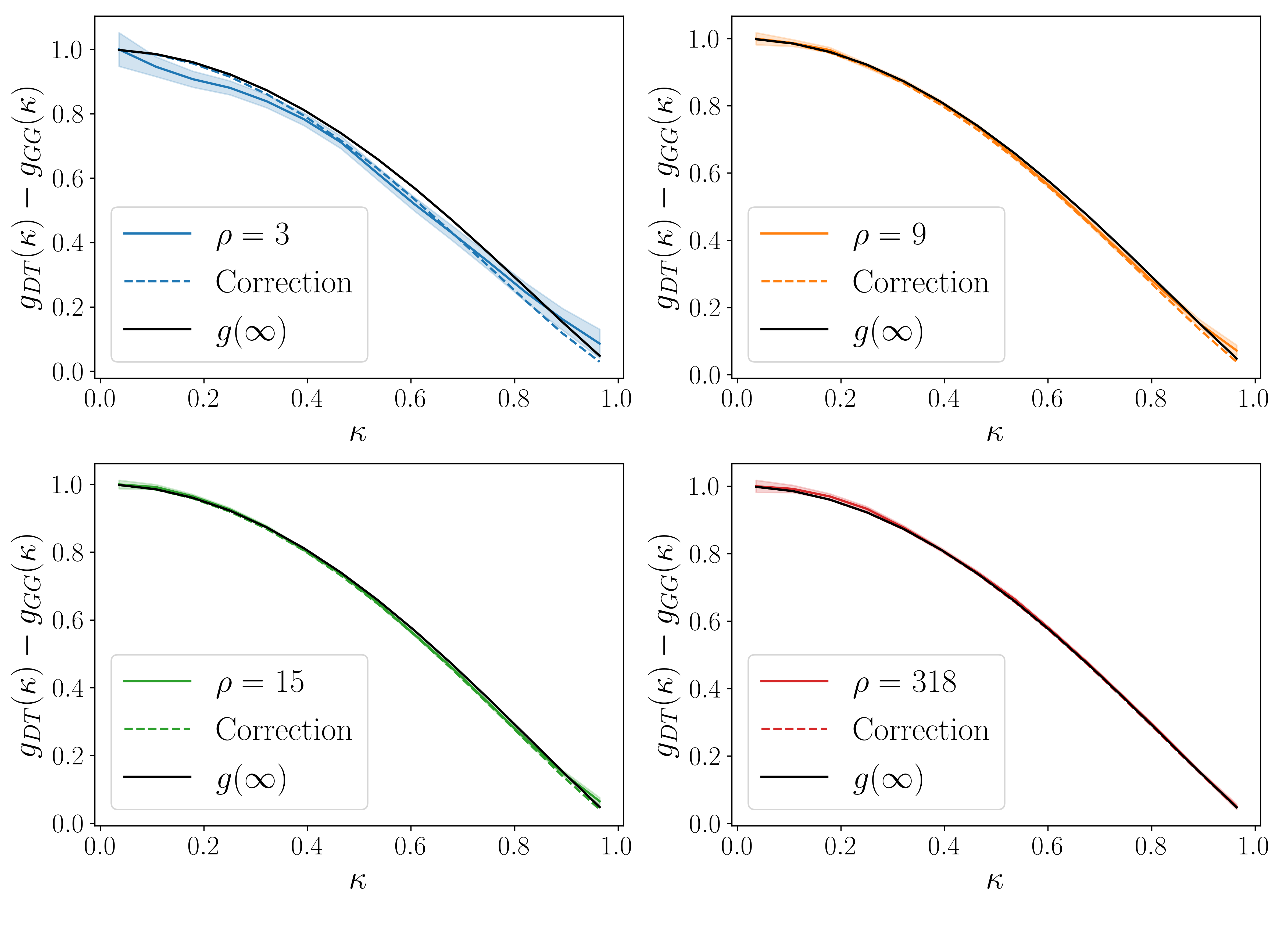}
\caption{Comparison of the expansion of Eq. \ref{eq:final} with the numerical result for the GG. The convergence to the infinite density limit is much faster than for other graphs and the approximation is very good for densities as low as $\rho=3$. The number of points is $N = \rho \pi$.}
\label{fig:gabriel}
\end{figure}

\begin{figure}
\centering
\includegraphics[width=0.5\textwidth]{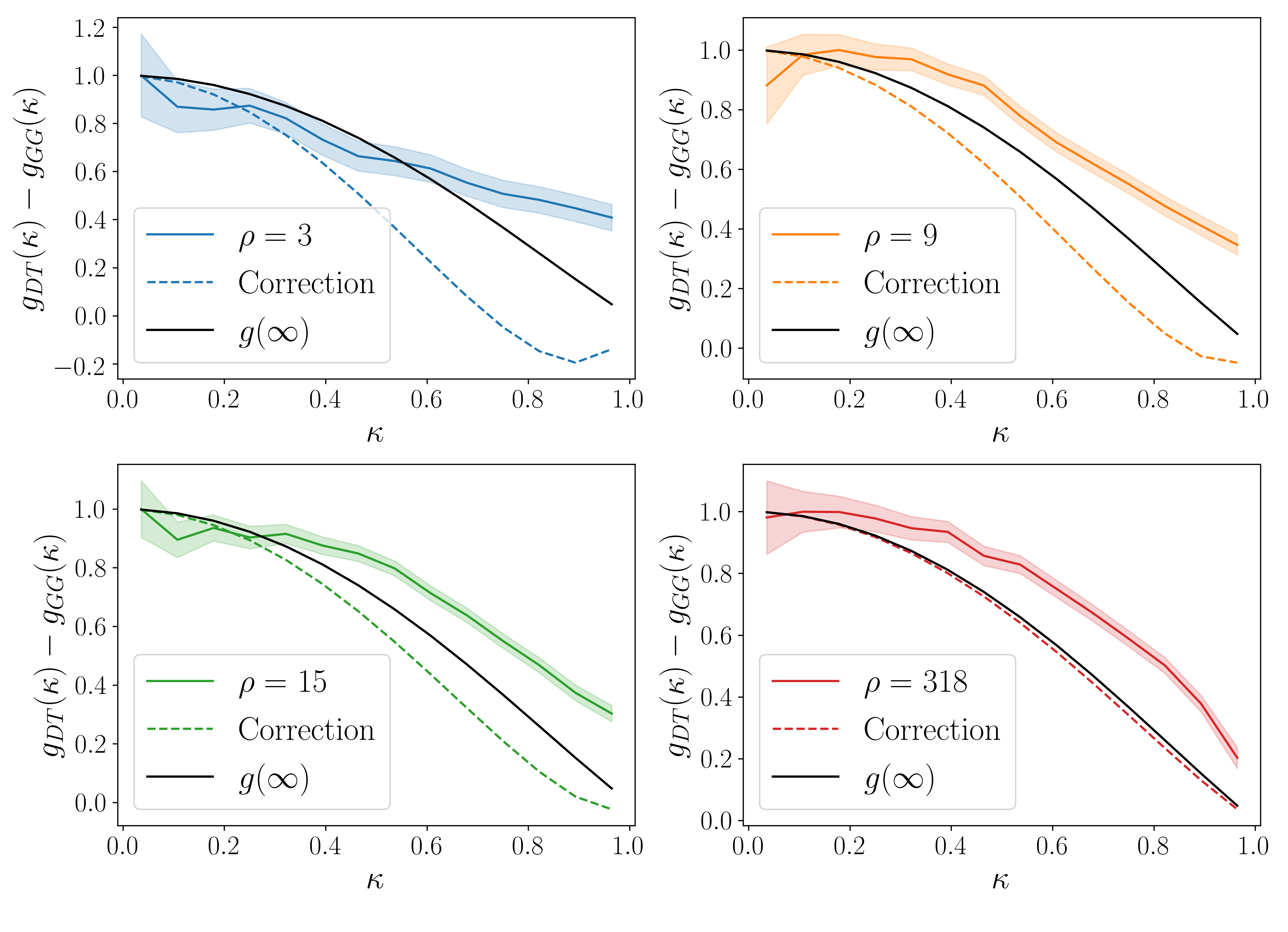}
\caption{Comparison of the expansion of Eq. \ref{eq:final} with the numerical result for the MST. The average BC converges towards the dense regime limit but the second-order approximation we used is not valid in this specific case.}
\label{fig:MST}
\end{figure}

We test the analytical result of Eq.~\ref{eq:final} on various graphs: the DT (Fig.~\ref{fig:DT}), the $k$-NN (Fig.~\ref{fig:kNN}), the RGG (Fig.~\ref{fig:RGG}), the GG (Fig.~\ref{fig:gabriel}) and the MST (Fig.~\ref{fig:MST}). For the numerical simulations, we sample $N$ random vertices in a disc (we test sizes from $N=10$ to $N=1000$) and connect the points according to the rules of each graph. We then compute the BC of each point using the Brandes algorithm \cite{Brandes:2004} and average the results over $5,000 runs$ (we build a new graph at each run).

For the GG, DT, $k$-NN and the RGG, we have an excellent agreement between the analytical result and our numerical simulations (averaged over 5~000 runs)  for $1 \ll \rho$.  The discrepancies between the quasi-dense and the dense regime are larger around $\frac{\kappa}{R}=0.8$. We observe that for these different graphs the speed of convergence to the infinite density limit is not the same. The convergence for the GG, RGG and the DT is fast while it is slower for the $k$-NN. For the GG, it is so fast that the infinite regime approximation is a good approximation for densities as low as 3 points per square unit (graphs of 10 points). In general, however, it is hard to predict or to understand why one type of graph would converge faster towards the infinite limit than the others. For $k$-NN and RGG graphs, the approximation is good from densities as low as 6 points per square unit (which corresponds to less than 20 points in the disc) while the approximation is valid for smaller densities (about 10 points in the disc) for DT and the Gabriel graphs (not shown here). We also note that Gabriel graphs (GG) being subgraphs of DT, we would naïvely expect that GG converge slower towards the dense regime limit than DT. This would result from the fact that the shortest paths are closer to straigth lines (and hence the dense regime) when more points are added in the network. This is not true however since the normalized expected BC of $\mathbf{\kappa}$ depends on both the average BC in $\mathbf{\kappa}$ and the maximal BC in the graph (in 0). Adding more points to the system can both decrease the BC on average but increase the maximal BC, leading to non-trivial behaviors of convergence towards the dense regime. 

Finally, we note that our approximation does not work for MST and RNG graphs since the assumption stating that $\chi(\epsilon)$ is independent of $(i,j, \kappa)$ seems not to be valid for these graphs. The BC however converges towards the infinitely dense limit result of \cite{Giles:2015} (see Fig. \ref{fig:MST} for the MST). At this point, it is an open question how to generalize our result in order to understand this behavior. 

\section{Note: 2d grid}

\begin{figure}
\centering
\includegraphics[width=0.4\textwidth]{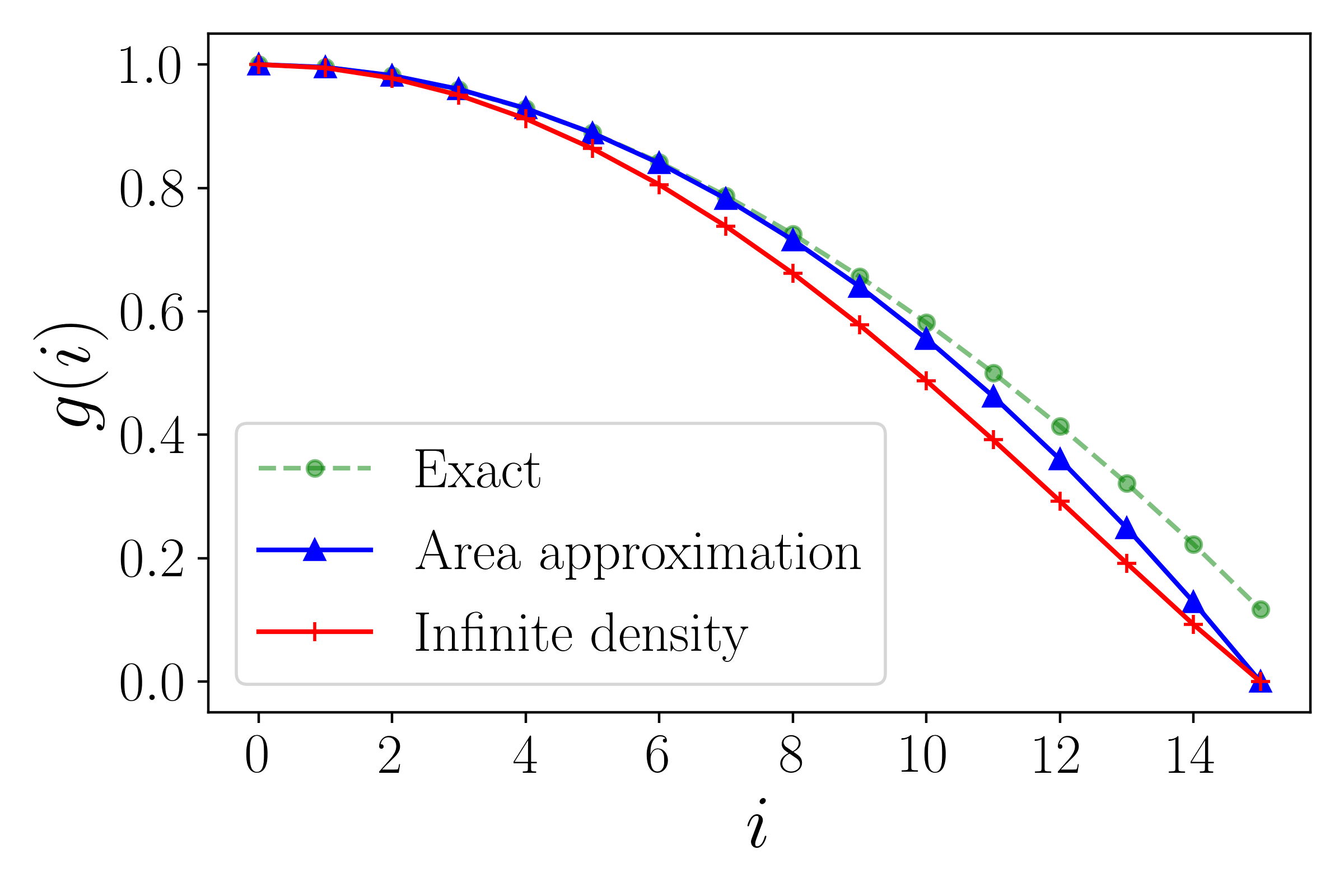}
\caption{Comparison of the exact numerical result for the 2d lattice, the approximation Eq.~\ref{eq:appr_2d}, and the result obtained for the infinite density Eq. \ref{eq:infinite}. We observe here that the infinite density result does not apply to this case. }
\label{fig:compare2d}
\end{figure}

We note that, somewhat surprisingly, for the 2d grid the above calculation doesn't apply. Indeed, we plot in Fig.~\ref{fig:compare2d} the exact numerical result for the 2d grid, the approximation Eq.~\ref{eq:appr_2d}, and the infinite density approximation (Eq. \ref{eq:infinite}). 
There are two main reasons why the infinite density calculation doesn't apply here. First, there is a strong degeneracy and the number of shortest paths is very large in general, and second, these shortest paths are not straight.  The main assumptions used in order to get the infinite density limit result \cite{Giles:2015} and our expansion do not therefore hold and we expect the observed discrepancy. However, the 2d grid case is not really a problem as we showed with the simple approximation  Eq.~\ref{eq:appr_2d}.

\section{Discussion}

In this paper we extended one of the few theoretical results about BC in spatial networks to a large number of families of graphs ($k$-NN, RGG, GG and DT) for non-infinite densities. We proved that for these families it is possible to find an approximation of the average BC of a random point in a bounded set of the plane just from its spatial coordinates. The infinite density limit which corresponds to the first term of our expansion is universal and independent from the graph. The first non-trivial correction encodes the deviations of shortest paths from the straight line and is therefore not universal. This approximation is theoretically valid for quasi-dense sets of points ($\rho \gg 1$) but is empirically correct for planar graphs with densities as low as a few points per square unit.

This approximation seems however not to be valid for other families of spatial networks (such as the RNG and the MST) whose dense limit is still universal but exhibit different convergence behaviors. The main difference comes from the way the shortest paths tend to straight lines and further studies are needed in order to understand this behavior. 

\begin{figure}
\centering
\includegraphics[width=0.5\textwidth]{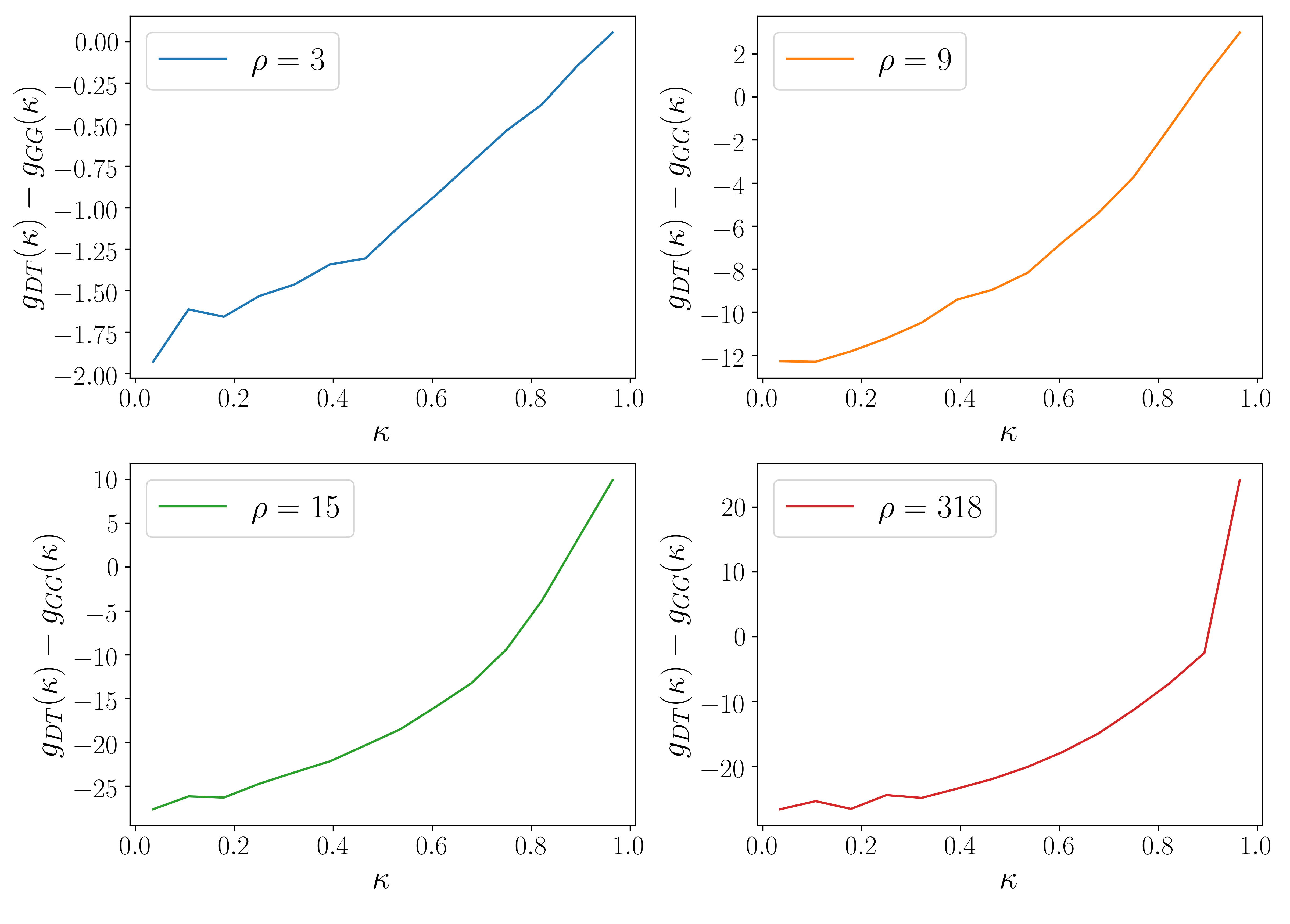}
\caption{On average, the BC decreases when constructing a Gabriel graph from a Delaunay triangulation. This results from the fact that the GG are subgraphs of DT. We note that the BC change is negative for all positions of $\mathbf{\kappa}$ except close to the edge of the disc, due to finite-size effects. These effects get more localized when increasing the density of points.}
\label{fig:DTvsGG}
\end{figure}

We also observed that adding more points to the network decreases the BC on average (see Fig. \ref{fig:DTvsGG}) as theoretically expected. However, locally, some points may be more central when new points enter the system and it is therefore not possible to predict the speed of convergence towards the dense-regime from just an inclusion relation: if $G_1\subset G_2$, there are more edges in $G_2$, but we can still have both a smaller average BC (due to theorem of \cite{Gago:2012}) and a larger maximal BC for this graph compared to $G_1$.  For example, the Gabriel graph is a subgraph of the DT and we observe that the difference $g_{DT}(\kappa)-g_{GG}(\kappa)$ has a sign that can be either positive or negative according to the value of $\kappa$ (Fig.~\ref{fig:DTvsGG}), implying that the convergence to the infinite density limit is not `uniform'. 

This theoretical work proposes a first step to study the BC in spatial networks but many questions are still open. As we mentioned, it is unclear why the behavior of the MST and the RNG is so different from the other graphs studied here. More work is certainly needed in order to understand how shortest paths in these systems become always more straight when the density increases. Also, an open question concerns the spatial patterns of the BC in disordered spatial networks and it would interesting to understand from a theoretical point of view the effect of disorder.

{\it Acknowledgements}. We warmly thank Alex Kartun-Giles for useful discussions and comments
at various stages of this work.  This material is based upon work supported
by the Complex Systems Institute of Paris Ile-de-France (ISC-PIF). VV
thanks Ars\`ene Pierrot (ISC-PIF) for his mathematical help and his general comments.

\bibliographystyle{prsty}

\end{document}